\documentclass[9pt,twocolumn,twoside]{article}
\pdfoutput=1

\RequirePackage{graphicx,xcolor}
\RequirePackage{colortbl}
\RequirePackage{booktabs}
\RequirePackage{tikz}
\RequirePackage{algorithm}
\RequirePackage[noend]{algpseudocode}
\RequirePackage{changepage}
\RequirePackage[left=48pt,%
                right=42pt,%
                top=48pt,%
                bottom=60pt,%
                headheight=15pt,%
                headsep=10pt,%
                letterpaper,twoside]{geometry}%
\RequirePackage[labelfont={bf,sf},%
                labelsep=space,%
                figurename=Figure]{caption}
\definecolor{black50}{gray}{0.5} % 50% black for hrules
\definecolor{color0}{RGB}{0,0,0} % Base color
\definecolor{color1}{RGB}{59,90,198} % unused
\usepackage{natbib} \bibpunct{(}{)}{;}{author-year}{}{,}
\RequirePackage[utf8]{inputenc}
\RequirePackage[english]{babel}
\RequirePackage{amsmath,amsfonts,amssymb}
\RequirePackage{mathpazo}
\RequirePackage[scaled]{helvet}
\RequirePackage[T1]{fontenc}
\RequirePackage{url}
\RequirePackage{hyperref}
\RequirePackage{lettrine}

\renewcommand{\d}{\mathrm{d}}

\title{Gene networks  accelerate evolution  by fitness
landscape learning}

\author{John Reinitz$^{\ast,1}$ \and
Sergey Vakulenko$^\dagger$ \and
Dmitri Grigoriev$^\ddagger$ \and
Andreas Weber$^\S$}
\date{$\ast$Dept.of Statistics, University of Chicago, Chicago, 57-Street,
USA,  \url{reinitz@galton.uchicago.edu} \\
$\dagger$Institute for Mechanical Engineering Problems, Russian Academy of Sciences,  Saint Petersburg, Russia
and  Saint Petersburg National Research University of Information Technologies, Mechanics and Optics,
Saint Petersburg, Russia, \url{vakulenfr@mail.ru}\\
$\ddagger$CNRS, Math\'ematiques, Universit\'e de Lille,
Villeneuve d'Ascq, 59655, France, \url{dmitry.grigoryev@math.univ-lille1.fr}\\
$\S$Dept. of Computer Science, University of Bonn, Bonn, Germany, \url{weber@cs.uni-bonn.de}}
\begin{document}
\maketitle

\begin{abstract}
We consider evolution of a large population, where fitness of each organism is defined by many phenotypical traits. These traits result from expression of many genes.  
We propose a new model of gene regulation, where gene  expression is controlled by  a gene network with a threshold mechanism  and there is a feedback between that threshold and  gene expression.   
We show that this regulation is very powerful: depending on parameters
 we can obtain any functional connection between thresholds and genes. 
Under general assumptions on  fitness we prove that such model organisms  are capable, to some extent, to  recognize the fitness landscape. 
That fitness landscape learning sharply reduces
the number of mutations necessary for adaptation and thus accelerates of evolution.  
Moreover, this learning increases phenotype robustness with respect to
mutations. However,  this acceleration leads to an additional risk since learning procedure can produce errors. 
Finally evolution acceleration reminds races on a  rugged highway:  when you speed up,  you have more chances to crash.  
These results  explain recent experimental data on anticipation of environment changes by some organisms. 
\end{abstract}

%\setboolean{displaycopyright}{true}

%\thispagestyle{firststyle}
%\marginmark
%\firstpagefootnote
%\correspondingauthoraffiliation{
%Dept.of Statistics, University of Chicago, Chicago, 57-Street,
%USA,  reinitz@galton.uchicago.edu
%}
%\vspace{-11pt}%

%\lettrine[lines=2]{\color{color2}T}{}his \textit{Genetics}
% journal template is provided to help you write your work in the correct journal format. Instructions %for use are provided below. 

%\section*{Guide to using this template in Overleaf}

%This template is provided to help you prepare your article for submission to the \textit{Genetics}.

%\section*{Author Affiliations}

%For the authors' names, indicate different affiliations with the symbols:
% $\ast$, $\dagger$, $%%\ddagger$, $\S$. After four authors, the symbols double, triple, %quadruple, and so forth as required.

\section{Introduction}
%\lettrine[lines=2]{\color{color2}T}{}he 
The central biological paradigm is that  evolution goes via gene mutations and selection.  This  process may be represented as 
a walk in a fitness landscape leading to a fitness increase and a slow adaptation \citep{Orr}.   According to  classical ideas this walk can be considered a sequence
of small random steps with small phenotypic effects. 
%\citep{neher2013genealogies}
However, in the 1980s, new experimental approaches
were developed, in particular, quantitative trait locus
 (QTL) analysis. In QTL analysis, the genetic basis of phenotypic
differences between populations or species can be
analyzed by  mapped molecular markers.
Genetic and molecular tools
allow us to find some genetic
changes that underlie  adaptation.  Results (see, for example, \citet{Zeyl})
show that evolution can
involve genetic changes of relatively large effect and, in some cases, the total number of changes are surprisingly small. Another intriguing fact is that
organisms are capable to make an 
adaptive prediction of environmental changes \citep{Nature}.
 
 To explain these surprising facts new evolutionary concepts were suggested (see the review
 by \citet{Watson} and references therein).   The main idea  is that 
 population can ``learn'' (recognize) fitness landscapes \citep{Vazirani, FV, Watson}. This learning can explain the adaptive prediction effect.

A mathematical basis for investigation of evolution learning problems  is developed by \citet{Valiant2006,Valiant}. However, this work uses a  simplified model, where
 organisms are represented as Boolean circuits  seeking for   an ``ideal answer'' on environmental challenges.   
 These circuits involve $n$ Boolean variables that can be interpreted as genes, and the ideal answer maximizes the fitness. 
 A similar model was studied numerically by \citet{FV} to confirm the theory of ``facilitated variation'' explaining appearance of genetic variations, 
 which can lead to large phenotypic ones. In the work by \citet{SatE}
an evolution theory of the Boolean circuits is advanced.  It is shown that, 
under some conditions---weak selection, see  \citet{Nagylaki})---a polynomially large population over polynomially
many generations (polynomial in $n$)  will end up almost surely consisting
exclusively of satisfying truth assignments. This
theorem can shed light on the problem of the evolution of complex
adaptations since that satisfiability problem can be considered as a 
rough  mathematical model of adaptation to  many constraints.

In \citep{Vazirani} it is shown that, in the regime of weak selection, population evolution 
can be described by the multiplicative weight update algorithm (MWUA), which is a powerful tool, well  known in theoretical computer science   
and generalizing such famous algorithms as Adaboost and  others \citep{MWU}.  Note that in \citep{Vazirani} infinitely large populations are investigated
whereas the results of \citep{SatE} hold only for finite populations and take into account genetic drift.

In this paper, we consider a  new model, which extends the previous ones and describes the Boolean circuits with genetic regulation.  By this model, we investigate a connection of the landscape learning problems with  evolution acceleration and    canalization.
The canalization effect,  pioneered
in the  paper by %Waddington  
\citet{Wad} means that the phenotype robustness becomes greater.  
 Canalization is a measure of the ability of a population to produce the same phenotype regardless of variability of its environment or genotype \citep{Wad}. 
 In our case it means robustness of adapted phenotypes with respect to mutations. 
 
Two seminal papers describe connections between evolution and canalization (phenotypic
buffering) quite differently.
 %Waddington 
 \cite{Wad}  claims that phenotypic buffering is needed for
adaptation. According to Waddington, there is a connection between canalization  and
genetic assimilation of an acquired character that was demonstrated experimentally by \citet{Wad1} for 
a population of Drosophila under a heat shock. 
However, based on other impressive experiments with Drosophila, %Rutherford and Lindquist 
\citet{RL}  say the exact opposite, namely, that
the  phenotypic buffering shutdown is required for adaptive changes. In the concept stated  by \citet{RL} and other works, for example
by \citet{Masel},   some genes can serve
as capacitors. When the population is under a stress (a heat shock, radiation, or chemical reagents),  buffering falls and capacitors release a hidden genetic information 
that leads to new species formation.  A mathematical model for this effect
is proposed in \citep{Biosystem}.

In this paper,  we aim to show that, in a fixed environment,
genes can serve as  learners.
We show, by analytical  methods, 
that the gene circuits having such regulation networks are capable to recognize, to some extent,  fitness landscapes. Indeed, if
an organism has survived within a long period, this fact brings an important information, which can be used for gene regulation system training.  Biological  interpretation of this fact is simple: if a population is large enough and mutations are sufficiently seldom, selection eliminates all negative mutants. So, 
if an organism is viable and it was affected by a mutation (which is not neutral, i.e. changed phenotype and the fithess), then with probability close to $1$ that mutation is positive.   
We obtain mathematical results, which allows us to estimate evolution acceleration and canalization due to that learning procedure.    
Note that the biological interpretation of evolution acceleration is also quite transparent: learning  by gene regulation networks sharply reduces 
the  number of mutations, which are necessary to form a  phenotypic trait useful for adaptation.

Note that we use a  model more sophisticated  than
the ones studied by \citet{SatE, Valiant}. In contrast to  these works, our model is not simply a Boolean
circuit.   
Namely,  Boolean circuits control  formation of quantitative phenotypic traits,  expressions of those phenotypic traits  range
in the whole interval $(0,1)$. However, in contrast to \citep{SatE, Valiant},
our circuits can be regulated, i.e., have a plasticity property.  
Note that biological ideas beyond that regulation mechanism were proposed, in a simpler and informal manner, still in  \citep{Stern}.  Stern introduced thresholds which determine how many
genes should be activated to express a trait useful for survival and 
explained mechanisms of gene assimilation suggested by %Waddington
\citet{Wad}. Actually, one can consider the model of  this paper as a combination 
of models    
\citep{SatE, Valiant, Stern}.  We show that fitness landscape learning is possible only if there exists a  a non trivial connection between a gene control of phenotype (morphogenesis)
and the gene regulation developed as a result of evolution.

\section{Materials and Methods}

In this section, we describe our model and our mathematical approach. 

\subsection{Genom} 

We assume that the genotype can be described by Boolean strings
\begin{equation}
s=(s_1, s_2,  \ldots , s_N) \label{gen1},  \quad  s_i \in S=\{0,1\}, \quad s \in S^N,
\end{equation}
i.e.,  a gene can either be  in an active state (switched on), or in a passive one (switched off).

\subsection{Phenotypic traits} \label{trait}

 Phenotypic traits are controlled by many genes \citep{Orr}.  We consider levels  $f_j$ of expressions 
of those phenotypic traits as real variables range in $(0,1)$.  
Then the vector $f=(f_1, \ldots ,f_m)$ can be considered formally as an organism phenotype.  We suppose that
\begin{equation} \label{Zj}
f_j=f_j(s, h_j), \quad j=1, \ldots , m,
\end{equation}
where $f_j \in [0,1]$ is a real valued function of the Boolean string $s$ (which is the genotype) 
and a real valued variable  $h_j$, which is a tuning parameter for gene regulation (we will describe 
it in more detail  below).

Only a part of $s_i$ involved in $f_j$.  Namely,  for each 
$j$ we have a set of indices $K_j=\{i_1, i_2, \ldots , i_{n_j}\}$ such that $f$ depends on $s_i$ only for $i \in K_j$, i.e.,
$$
f_j(s, h)=f_j(s_{i_1}, s_{i_2}, \ldots , s_{i_{n_j}}, h_j)
$$
  where $i_l \in K_j$ and $n_j$ is the number of elements in $K_j$ ($n_j$ is the number of genes involved in the control of  the trait expression).

Another possible interpretation of $f_i$ is as follows. Multicellular organisms consist of cells of different types. One can suppose that the organism phenotype is defined completely by the corresponding cell pattern. 
The cell type $j$  is determined by  morphogenes, which can be identified as  gene products 
or  chemical reagents that can change cell type (or  genes that code for that chemical reagents that can determine cell types or cell-cell interactions
and then finally the cell pattern). The morphogene activity is defined by (\ref{Zj}).

We suppose that the following assumption is satisfied:

\vspace{0.2cm}

{\bf Assumption M}. {\em Assume that activities $f_j(s, h)$ have the following properties.

{\bf M1} 
All $f_j(s, h_j)$ are  functions of real valued parameter 
$h_j$.    For each  fixed $s$ 
\begin{equation} \label{hdec}
 f_j(s, h_j) \to 0 \ (h_j \to - \infty), 
\end{equation}
 and
 \begin{equation} \label{hdec2}
  f_j(s,h_j) \to 1 \  (h_j \to +\infty)
\end{equation}

{\bf M2} 
 The sets $K_j$ are independent random subsets of  the set of all genes consisting at most $K$ elements:
\begin{equation} \label{Kj}
K_j=\{i_1, \ldots, i_{n_j}\},   \quad i_l \in  \{1, \ldots , N\},   \quad 0 < n_j \le K,   
\end{equation}
and 
\begin{equation} \label{free}
 Km/N=\alpha  < 1.
\end{equation}
} 
\vspace{0.2cm}

The second assumption entails that only a part of all genes  is involved in control of the phenotypic expression.  We denote the number of genes involved in regulation of $f_j$ by $N_r$.  
Note that  (\ref{free}) shows that  $N_r=\sum n_j \le Km$.

%\vspace{0.2cm}
\medskip\noindent

Condition {\bf M1} implies that there exists a parameter in $f_j$, which can control this  function.   Assumption {\bf M2} means
that the genes are organized, in a sense, randomly (note that this assumption plays a key role in probabilistic methods for
problems with many constraints such as $k$-SAT \citep{Friedgut}). The condition $\alpha  \ll  1$ means that we have a "gene freedom", i.e., we have sufficiently many genes
with respect to the number of the traits. This freedom yields that the probability of gene pleiotropy in the gene regulation is sufficiently small for large genomes. 

Let us consider a biological interpretation of $h_j$.  We consider $h_j$ as a tuning parameter that defines
 level and sign of gene regulation for $j$-th phenotypocal trait.  The genes produce a number of different gene products (proteins, miRNA,
tRNA). Some of these products can serve as transciption factors involved in gene expression regulation.
The parameters $h_j$ determine  intensities and signs of this regulation. Condition (\ref{hdec}) and (\ref{hdec2}) mean that
for large negative $h$ one has a strong repression of  $f_j$ expression while for large positive $h$  the expression level is close to almost maximal one.

Consider a biologically natural  example, where  the assumptions {\bf M1} and {\bf M2} hold. This 
example is inspired by the work \citep{Stern} and model \citep{Mjol}. Let  
\begin{equation} \label{cellj}
f_j = z_j=\sigma(\sum_{i=1}^N w_{ji}{s_i} -  h_j),
\end{equation}
where $j = 1, \ldots, m$.  
Here $\sigma(S)$ is a sigmoidal function of $S$ such that 
\begin{equation} \label{sigma}
\sigma(+\infty)=1, \quad \sigma(-\infty)=0, \quad  \sigma^{'}(S) > 0.
\end{equation}
As an  example, we can take $\sigma(S) =(1+\exp(-b S))^{-1}$, where $b>0$ is a sharpness parameter. Note that for large $b$ 
this sigmoidal function tends to the step function and for $b=+\infty$ our model becomes a Boolean one. 
Some experimental results show that miRNA control involves a threshold mechanism \citep{RNA}. 
Note 
that threshold mechanisms  are  omnipresent in gene networks \citep{Kauff2,Mjol} and fundamental for neural networks \citep{Hop}. 
The  parameters $h_j$ are important tools of gene regulation since they affect  the gene circuit structure and define circuit plasticity. 
The idea to use such a parameter was proposed by \citet{Stern}, see also
\citep{SatE} for interesting comments on the connection with evolution.    

Let us introduce the matrix $W$ of size $m \times N$ with  entries $w_{ij}$. 

The coefficients $w_{ji}$  determine the effects of  terminal differentiation genes  \citep{Biosystem}, and hence
encodes the genotype-phenotype map. We assume that $w_{ji} \in \{[-1- \gamma, -1], 0, 
[1, 1+\gamma]\}$, where $\gamma$ is a positive parameter.  Moreover, we assume that
the coefficients $w_{ji}$ are random, with the probability that $w_{ji} >0$  or that  $w_{ji} < 0$
is $\beta/2N$, where $\beta >0$ is a parameter. 
This quantity $\beta$ defines a genetic redundancy, i.e., averaged numbers of genes involved
in regulation of a   trait. Note that then assumption {\bf M2} holds and $K \approx \beta$ for large $\beta$.

 The number $\gamma$ is a measure of
phenotypic buffering. 
The condition  $\gamma= 0$ means that there is no phenotypic buffering, 
 $\gamma > 0$ means
that the buffering works; larger 
 $\gamma$ leads to a buffering increase.

 Let us introduce the matrix $W$ of size $m \times N$ with the entries $w_{ji}$ and the corresponding
 the sign matrix $S(W)$ with entries $s_{ij}=sign(w_{ij})$, where $sign(x)=1$ for
 $x > 0$, $sign(x) =-1$ for $x < 0$ and $sign(0)=0$. The matrix  $S(W)$ plays an important role below;
 that matrix determines qualitatively the control of the phenotypical traits by the genes. If $s_{ji} >0$ this means that activation of $i$ -th gene leads to an increase of $j$-th trait expression; if $s_{ji} < 0$ then activation of $i$ -th gene leads to a decrease of $j$-th trait expression; at last, if $s_{ji}=0$ this means
 that the $i$-th gene is not involved in the control of $f_j$.

\subsection{Fitness}

Actually we know a little about fitness of multicellular organisms, see e.g. the review by \citep{Krug}.  Recall 
some known fitness models.  

The  random field models assign fitness values to genotypes independently
from a fixed probability distribution. They are close to 
mutation selection models introduced by \citet{King}, and can be named 
House of Cards (HoC) model.
The best known model of this kind is 
 the NK model introduced by %Kauffman and Weinberger 
 \citet{Kauff2}, where each locus interacts
with $K$ other loci. 
Rough Mount Fuji (RMF) models are obtained by combining a random HoC
landscape with an additive landscape models \citep{Aito}.

In this work, we use the classical approach of R. Fisher, namely, phenotype-fitness maps. Our phenotype is given by $f$,
i.e., we assume that the phenotype is completely determined by the phenotype trait expression, and thus
the fitness depends on $s$ via $f_j$.
 
We can express  the relative fitness via an auxiliary function $W(s)$  by relation
\begin{equation} \label{Fi}
F=C_M \exp(W(s)),  
\end{equation}
where a constant $C_M$ is proportional to  the number of progenies. 
Below we refer $W$ as  a fitness potential, and we assume that
\begin{equation} \label{envW}
W= \sum_{j=1}^{L}  b_j   f_j(s, h_j).  
\end{equation}

We consider fitness as a numerical measure  of interactions between
the phenotype and an environment. For a fixed environment, this
idea gives us the  fitness  of classical population genetics.
A part of the fitness, however, depends on the organism developing properly
and for now we represent it as independent of the environment (we are aware
that this is not always true).

A part of coefficients $b_j$ may be negative and the other part is positive. The corresponding contributions will be denoted by $W^I$ and $W^E$, respectively:
\begin{equation} \label{envWp}
W^I= \sum_{j=1, b_j < 0}^{L}  b_j   f_j(s, h_j),  
\end{equation}
\begin{equation} \label{envWm}
W^E= \sum_{j=1, b_j >0}^{L}  b_j   f_j(s, h_j).  
\end{equation}

The first function $W^I(s)$  is associated with the internal fitness and the second $W^E(s)$ determines an interaction between the organism and its environment. 
If we assume that $f_j$ are morphogene concentration levels, which control cell types, then the part  $W^I$  measures a viable development in
terms of  formation of correct cell types. Each cell type is determined by the corresponding morphogene activity $f_j$. Another component of fitness,
$W^E(s)$
 depends on the environment and it describes how well the organism is adapted to
it. 
 The terms $ f_i$ involved in $W^E$ can be interpreted as gene responses on the environment.
The terms with $b_i <0$ involved in $W^I$  can for example define a fitness reduction caused by  formation of non-necessary (excess) cells.  We assume that in a ``normal'' state, that corresponds to
the maximal fitness,  all $f_j$ in $W^I$ are close to zero. When such a $ f_j \approx 1$, this can be interpreted as appearance of a "bad" cell, for example, a cancer one. 

Another possible interpretation of $W^I$ is that a larger  expression of some phenotypical traits
can decrease chances of the organism to be viable in a given environment.

 Note that
this model (\ref{envW})   can describe gene epistatic effects  via dependence of $f_j$ on $s$ if $f_j$ are nonlinear in $s$.

\subsection{Population dynamics model} \label{popdyn}

For simplicity, mainly we consider  populations with asexual reproduction (although a part of results
is valid for sexual reproduction, see comments in the end of this subsection).

In each generation , there are $N_{\mathrm{pop}}(t)$ individuals,
the genome of each of them is denoted by $s(t)$, where $t=0,1,2,\ldots $ stands for the evolution step number).
 Following the classical ideas of Wright -Fisher model, we suppose that
generations do not overlap. In each generation (i.e., for each $t$), the following three steps are
performed:

\begin{enumerate}
\item Each individual $s$ at each evolution step can mutate  
 with probability $p_{\mathrm{mut}}$.

\item At  evolution step $t$ each individual produces $k$ progenies randomly with the probability $P_k$ defined by the  Poisson law
\begin{equation} \label{Poisson}
P_k=\frac{q(t)^k}{k!} \exp(-q(t))=\Lambda(k, q(t)),
\end{equation}
where $q=F/\bar F(t)$, $F$ is the fitness of that individual and $
\bar F(t)$ is the averaged fitness of the population at the moment $t$ defined by. 
\begin{equation} \label{popdyn2} 
\bar F(t)=\sum_{s  \in S(t)} X(s, t) F(s(t)), 
\end{equation}
where $\bar F(t)$ can be interpreted as the averaged population fitness at the moment $t$ , $S(t)$ the set of the genotypes 
represented in the population at the moment $t$ (the genetic pool) and
$X(s,t)=N(s,t)/N_{pop}(t)$ is the frequency of genotype $s$. Here $N(s,t)$ denotes the number of the population members with 
genotype $s$ at the step $t$.

\item If $N_{\mathrm{pop}}(t)> N_{\mathrm{popmax}}$, where $N_{\mathrm{popmax}}$ is some number,  individuals are removed randomly until  $N = N_{\mathrm{popmax}}$.
\end{enumerate}

After each selection step, there occur mutations in the genotypes, which create a new genetic pool  and then a new round of selection starts.   
The last condition express the fundamental ecological restriction that all environments have restricted  resources only (bounded capacity), 
 therefore, they can supply only populations bounded in size.  
However, if the evolution time $T$ is bounded, $t =0,1, \ldots T$ and $N_{pop}(t) \gg 1$ then we can remove the last condition 
since by (\ref{Poisson}) and the Central Limit Theorem one can show
that fluctuations of the population size are small: $|N_{\mathrm{pop}}(t+1)-N_{pop}(t)| \ll N_{\mathrm{pop}}(t)$.  
Thus then the population is ecologically stable and the population size fluctuate weakly.  The condition (3)
is really essential for small populations only (which considered by numerical simulations).

In the limit case of infinitely large populations we use the following dynamical  equations for the frequency $X(s, t)$  of the genotype $s$ in the population at the moment $t$:
\begin{equation} \label{popdyn1} 
X(s, t+1)= \bar F(t)^{-1}  X(s, t) F(s(t)),   \quad X(s, 0)=X_0(s).
\end{equation}

Equations (\ref{popdyn1}) do not take into account the genetic drift. For large but finite populations $N_{\mathrm{popmax}}$ 
we should take into account this effect. Moreover, it is important that our populations and organisms
can extinct.

Equations (\ref{popdyn1}) and  (\ref{popdyn2}) describe a change of the genotype frequencies  due to selection at the $t$-th evolution step.  
The same equations govern evolution in the case of sexual reproduction in the limit of weak selection
\citep{Nagylaki, Vazirani}. Note that for an evolution defined by (\ref{popdyn2}), (\ref{popdyn1}) the averaged fitness $\bar F(t)$ satisfies  Fisher's theorem, 
namely, this function does not decrease in evolution step  $t$ and
we have $\bar F(t+1) \ge \bar F(t)$.

Note that  for simplicity we consider the point mutations (see the point {(i)} above) although it is well known that
mutation process is much more complicated. However, some of our  analytical results are valid for more general situations.

\subsection{Gene regulation} \label{GR}

We introduce the regulatory genes $y_j$, where $j=1,\dots, n$ and $n$ is the number of regulatory genes. They may be hubs in the networks, i.e., interact with many  genes.
The activities of $y_j$ are real numbers  defined by
\begin{equation} \label{yH}
y_j(s)=\sigma_R( a_{j1} s_{1}+  \ldots  + a_{jN} s_{N} - \tilde h_j), 
\end{equation}
where $\sigma_R$ is a sigmoidal function (one can take here a linear approximation, for example,  $\sigma_R(x)= a_R x$, where $a_R$ is a coefficient ), 
 $a_{ji}$  are real valued coefficients and $\tilde h_j$ are thresholds.

The feedback between genotype   $s$ and the parameters $h_j$ is defined by a dependence of thresholds $ h_j$ via the quantities 
$y_j$'s:
\begin{equation} \label{barhy}
h_j(s) =\bar h_j -   r_F \sum_{\tau=0}^{\tau_r} \sum_{k=1}^n  g_{jk}(\tau) y_k(s(t-\tau)),
\end{equation}
where $g_{ik}$ are positive random numbers, $r_F >0$ is a feedback parameter and $\bar h_j$ are positive constants. The parameter $\tau_r$ defines the memory of the regulation network.

If we introduce the shift map 
  $T_{\tau}$ defined  on sequences $s(t)$ by $T_{\tau}s(t)=s(t-\tau)$, then $h_j$ becomes a function
of shifted Boolean arguments 
\begin{equation} \label{HH}
h_j=h_j(s(t), s(t)^{(1)}),\ldots , s(t)^{(\tau_r)}). 
\end{equation}
where $s^{(\tau)})=T_{\tau} s(t)=s(t-\tau)$.

\subsection{Main assumptions}

%{\bf A}. We assume that at the initial evolution moment $t=0$ each genotype $s$ is represented
%in a sufficiently large number of organisms, i.e. the number $N(s)$ of beings with the genotype $s$
%satisfies
%\begin{equation} \label{freqgen}
%N(s) \ge N_{\mathrm{pop}}^{\mu},  \quad \mu \in (0,1).
%\end{equation}

{\bf A}. We assume that the mutation probability $p_{\mathrm{mut}}$ is small and  the time $T$ of evolution is large:
\begin{equation} \label{freq}
\theta=N p_{\mathrm{mut}}   \ll  1,   \quad T  \gg  \log(N_{\mathrm{popmax}}).
\end{equation}

%{\bf C}.  Let us suppose that the population abundance $N_{\mathrm{pop}}$ is large enough and $N_{\mathrm{pop}} \theta  \gg  1$, however, $\log N_{\mathrm{popmax}}  \ll  T$.

{\bf B}.  We choose initial genotypes randomly from a gene pool and assign them to organism. 
This choice is invariant with respect to the population member, i.e,. the probability 
to assign a given genotype $s$ for a member does not depend on that member. 

%The assumption {\bf  B} allows us to overcome computational difficulties outlined above. Indeed, this condition
%guarantees that  the population does not extinct even if the fitness does not attain its maximal value. 

%The condition {\bf D} is critically important. Only large populations are capable 
%to recognize the fitness landscape correctly. 

\subsection{Complexity of the model}

\subsubsection{Connections with hard combinatorial problems}

Adaptation (i.e., maximization of the fitness) is a very hard problem, since in evolution history we observe
 coevolution of many traits. 
As an example of such coevolution, we can consider mammal evolution. Long evolution of mammals is marked by development of many traits.  
Mammals are noted for their large brain size relative to body, size, compared to other animal groups, moreover, 
mammals developed many other features: lactation, hair and fur, erect limbs, warm bloodness etc. 
It is not clear how a random gradual search based on  small random mutation steps and selection only, can resolve such complex adaptation problem 
with many constraints and to create a complex phenotype with many features. 

To show that the model stated above reflects this biological reality, let us consider the case, where $f_j$ are defined by relations  (\ref{cellj}) and assume that

{\bf i}) 
$\sigma$ is the step function; 

{\bf ii}) $b_j >0$.

As a consequence of the second assumption   $F$ attains its maximum for $f_1=1, f_2=1, \ldots , f_m=1$.
Let us show that, even in this particular case,  
the problem of the fitness maximization with respect to $s$ is very complex.  In fact, 
for a  choice of $h_j$ it  reduces to the famous NP-complete problem, so-called $k$-SAT, which has been received a great attention of mathematicians, computer scientists, and biologists the last decades
(see \citep{Cook, Levin, Friedgut, Moore}).  The $k$-SAT can be formulated as follows.

{\bf $k$-SAT problem}
{\em
Let us consider the set
$V_n=\{s_1, \ldots , s_n\}$ of Boolean variables $s_i \in \{0,1 \}$ and a
set ${\mathcal C}_m$ of $m$ clauses. The clauses $C_j$ are
disjunctions (logical ORs) involving $k$ literals $z_{i_1}, z_{i_2},
\ldots , z_{i_k}$, where each $z_i$ is either $s_i$ or the negation $\bar
s_i$ of $s_i$.  The problem is to test whether one can satisfy all of the
clauses by an assignment of Boolean variables}.

%It can be illustrated by the following picture (see Fig.\ \ref{Fig1}).

%\begin{figure}[h!]
%\vskip-0.5truecm
% \includegraphics[width=80mm]{qwerty3.jpg}
%\caption{\small This image illustrates a toy $k$-SAT problem for $n=4$ logical variables $s_1, s_2, s_3, s_4$ for $k=3$  and $m=2$. 
% The clauses are  triangles, here
% we have  the two clauses of length $3$.  
% An  assignment of the logical variables is shown by arrows.
%This assignment is correct and gives
% a solution of the problem  if all the clauses are true.  In this case  we can take, for example,  $s_1=0,  s_2=0, s_3=1$ and $s_4=1$.
%This toy example is easy to resolve, however, the problem becomes difficult for $n , m \gg 1$. In our biological model, we interpret
%$s_i$ as genes  and clauses as  gene expression patterns for different cell types.}\label{Fig1}
%\end{figure}

Cook and Levin \citeyear{Cook, Levin} have shown that  $k$-SAT  problem is NP-complete and therefore in general it is not feasible in a reasonable running time.  
In subsequent studies---for instance, by \citet{Friedgut}---it was shown that $k$-SAT of a random structure is feasible 
under the condition that $m < 2^k N$. 
%In our case this means that the genetic redundancy is large enough.

To see a connection with $k$-SAT,  consider relation (\ref{cellj})  under assumption $w_{ij}=1$ 
and set $ h_j=-C_j+0.5$, where $C_j$ is the number of negative $w_{ji}$ in the sum $S_j=\sum_{i=1}^N
w_{ji}{s_i}$.
Under such choice of $h_j$ terms $\sigma(S_j)$ can be represented as disjunctions of literals  $z_j$.
Each literal $z_j$ equal either $s_j$ or  $\bar s_j$, where $\bar s_j$ denotes negation of $s_j$.  To maximize the fitness  we must assign such $s_j$ that all 
disjunctions will be satisfied.  If we fix the number $k$ of the literals participating in each disjunction
(clause),  this assignment problem is $k$-SAT  
  formulated above.

\subsubsection{ Biological interpretation of $k$-SAT}

Reduction to the $k$-SAT is a  transparent way of
representing the idea that multiple constraints need to be satisfied.  
The quantity $k=\beta$ define the gene redundancy and the probability of gene pleiotropy. For larger 
$\beta$ this probability is smaller.
The threshold parameter $h_j$ and $\beta$ define the  number  of genes, which 
 should be flipped to attain a need expression level of the trait  $f_j$. Mathematical results
mentioned above say us that for a fixed $N$ and $m$ constraints can be satisfied  for sufficiently large
$\beta >\beta_0(N,m)$ only.

Note that there are important differences between $k$-SAT in Theoretical Computer Science
and the fitness maximization problem.  First,  the signs of $b_j$ are unknown for real biological situations
since the fitness landscape is unknown.  The second, our adaptation problem involves the threshold 
parameters $h_j$ (see (\ref{cellj})).   In contrast to Computer Science Theory,  in our case the Boolean circuit has a plasticity, i.e.,  $h_j$ are not fixed.

If $b_j$ are unknown, the adaptation (the fitness maximization) problem becomes even  harder because
we do not know the function to optimize.  Therefore, many  algorithms for $k$-SAT are useless for biological adaptation problems. Below nonetheless  we will obtain some analytical results under 
assumption that $b_j$ are random.

\subsubsection{ How to accelerate evolution? Main ideas}

To overcome the computational hardness of our model, we apply the following ideas.  By assumption {\bf M2} we use the randomness of 
the gene organization of  expression and a small probability of gene pleiotropy.  Moreover, for an organism survival  it is not obligatory to attain the global maximum of the fitness, 
it is enough to attain a fitness value which is greater than fitnesses of other competing organisms.

However, the key idea inspired by the paper of  \citet{Stern} is as follows.  Suppose a fitness landscape learning is possible and the signs of $b_j$ become known as a result of evolution
(we will describe in the next sections  how that learning can work). Let, for example, 
$b_{10} >0$.  Then one can use circuit plasticity, i.e., a possibility to change $h_{10}$.  
In $k$-SAT, where $h_{j}$ are fixed,  we seek for correct values $s_i$ involved in the right hand sides of (\ref{cellj}) and it may be a computationally hard  problem. In our case we just strongly increase or decrease of $h_j$, depending on the sign of $b_j$. It can be done by a gene regulation loop, described
in  subsection \ref{GR} but only under a correct choice the gene regulation scheme and parameters. 

\section{Results}

The main results can be outlined as follows.
Here we first use ideas analogous to ones given by \citet{Vazirani} and also we propose an alternative method based on R. Fisher's
theorem.    
The second  approach allows us to see for which fitness functions the fitness landscape learning 
is possible,  to find optimal regulatory mechanisms and to investigate how this regulation depends on the  fitness function structure. The gene regulation rate should be smaller for more rugged fitness functions.

\subsection{Gene Regulation Power}

The gene regulation defined by relations (\ref{yH}) and (\ref{barhy}) is very powerful. It follows from 
the next assertion.

{\bf Theorem I } \label{T1} 

{\em Let $H_1(s, s^{(1)}, \ldots, s^{(\tau_r)})$, \ldots, 
$H_{m}(s, s^{(1)}, \ldots, s^{(\tau_r)})$  be functions of $\tau_r+1$ independent Boolean arguments
$s, s^{(1)}, \ldots,  s^{(\tau_r)}$.  Then for any $\epsilon >0$ there exist parameters $a_{kl}, \tilde h_k, g_{ik}(0), \ldots, g_{ik}(\tau_r)$
in (\ref{yH}) and (\ref{barhy}) such that $h$ defined by (\ref{HH}) satisfies
$$
|h_j(s, s^{(1)}, \ldots, s^{(\tau_r)}) - H_j(s, s^{(1)}, \ldots, s^{(\tau_r)}| <\epsilon 
$$
for all $j$ and $s$.
}

Roughly speaking  this means that gene regulation networks  defined by  (\ref{yH}) and (\ref{barhy})  can approximate with arbitrary accuracy 
any prescribed time delayed  feedback. 

{\bf Proof}.   The theorem follows from approximation results for multilayered perceptrons.  In fact, combination
of (\ref{yH}) and (\ref{barhy}) defines a straight forward neural network, namely, a perceptron with a single hidden layer.  
It is well known that such two layered perceptrons can approximate any Boolean target functions;
for a proof see \citep{Barron}.

\subsection{Fitness Landscape Recognition Theorems}

The following results are based on ideas close to ones given by \citet{MWU} and \citet{Vazirani},
but, for simplicity, we consider asexual reproduction. 
To obtain similar results for sexual reproduction, one can consider a weak selection regime 
and use classical results of \citet{Nagylaki}. 

Let us introduce two sets of indices $I_{+}$ and $I_{-}$ (that we refer in sequel as  positive sets and  negative ones, respectively)
such that $I_{+} \cup I_{-}=\{1, \ldots , m \}$.  We have  
\begin{eqnarray} \label{delta2}
I_{+}=\{ j \in \{1, \ldots , m \}| b_j >0 \}, 
\\
 \label{delta2m}
 I_{-}=\{ j \in \{1, \ldots , m \}| b_j < 0 \}.
\end{eqnarray}
The biological interpretation of that definition is transparent: expression of the traits $f_j$ from the positive set $I_{+}$  increases the fitness. For 
the negative set $I_{-}$   that expression decreases the fitness. 

Moreover, let us introduce  useful auxiliary sets.  Let $s$ and $s'$ be two genotypes.  Then we denote by
$Diff(s, s^{'})$   the set of positions $i$ such that $s_i \ne s_i^{'}$:
$$
 Diff(s, s') =\{ i \in \{1,\ldots , N\}|  \  s_i \ne s_i^{`} \}.
$$
That set contains gene positions of  the Boolean genome, where genes are flipped. 
 
We also use notation
$$
S_b= \sum_{j=1}^m b_{j, +},  \quad  b_{j, +}=\max\{b_j, 0\}.
$$
Note that
 
\begin{equation} \label{SB}
\max_{s \in \Pi^N}   F(s) \le S_b.
\end{equation}

Indeed,  the fitness $F(s)$ attains the global 
maximum when $f_j=0$ for all $b_j  < 0$ and $f_j=1$ for all $b_j >0$.
 
Below we prove two theorems on fitness landscape learning. 
First we consider the case of
infinitely large populations. 

{\bf Evolution Recognition Theorem II.} \label{T3}

{\em  Suppose  that evolution of genotype frequencies  $X(s,t)$ is determined by equations (\ref{popdyn2}) and (\ref{popdyn1}). Moreover, assume that

{\bf I}  for all $t \in [T_1, T+ T_c]$, where $T_1, T_c >0$ the population contains two genotypes $s$ and $s'$ such that 
the frequencies $X(s, t)$ and $X(s', t)$ satisfy
\begin{eqnarray} \label{zzz1}
X(s,T_1) &= &p_0>0,  
\\
 \label{zzz1a}
 X(s', t) & > &p_1 >0 \quad \forall t \in [T_1, T_1 +T_c];
\end{eqnarray}

{\bf II}  we have
\begin{equation} \label{Diff}
Diff(s, s^{`} ) \subset K_j,  
\end{equation}
for some $j$, i.e., the genes   $s_i$ such that $s_i \ne s_i^{`}$ are involved in a single regulation set $K_j$; moreover,
$$
\delta=|f_j(s, h_j) -f_j(s', h_j)| >0;
$$

{\bf III}  
\begin{equation} \label{kappa}
\kappa= min_{j} |b_j| >0,   
\end{equation}
 and all $h_j$ are fixed on the interval $[T_1, T_1 +T_c]$.
 
 Let  $T_c$ satisfy
 \begin{equation} \label{zzz}
T_c > \frac{-  \log(p_0 p_1)}{1 +\varepsilon},  \quad \varepsilon= \frac{\kappa \delta}{S_b},   
\end{equation}
 Then, if  
\begin{equation} \label{zincr}
f_j(s, h_j) <  f_j(s', h_j),
\end{equation}
we have  $ j \in I_{+}$.  If $f_j(s, h_j) >  f_j(s', h_j)$,   then $j \in I_{-}$.
}

Before proving let us make some comments. The biological meaning of the theorem is very simple:

{\em  For fitness models, where unknown parameters $b_j$ are involved in a linear way, absence of pleiotropy in gene control of  phenotypic traits $f_j$  lead to the fitness landscape learning in the limit
of infinitely large populations. 
}

Moreoever, let observe that we do not make no specific assumptions to the mutation nature, they may be point ones or more complicated but it is important that all gene variations  between  $s$ and $s'$
are contained in  a single regulatory set $K_j$.   

{\bf Proof.}  The main idea beyond the proof is very simple. Since
genetic drift is absent, the negative
mutations leads to an elimination of mutant  genotype from the population, the corresponding frequency becomes, for large times, exponentially small.

Consider the case (\ref{zincr}). Let $j \in I_{-}$. Consider the quantity $Q(t)=\frac{X(s, t)}{X(s', t)}$.  We observe that
\begin{equation} \label{diffF}
\Delta F=F(s) -  F(s')=b_j (f_j(s, h_j) -f(s', h_j)).
\end{equation}
Note that if $j \in I_{-}$ then assumptions {\bf II} and {\bf III}
 entail that
\begin{equation} \label{diff}
\Delta F \ge \kappa \delta.
\end{equation}

Relation (\ref{diff}) implies
$$
\frac{F(s)}{F(s')} = 1 + \frac{\Delta  F}{F(s')} \ge 1+ \frac{\Delta  F}{\max F(s')}.
$$
Due to (\ref{SB}) one has
$$
\frac{F(s)}{F(s')} \ge 1+ \varepsilon.
$$
According to (\ref{popdyn1})  the last inequality implies that for $T > T_1$
\begin{equation} \label{QQ}
Q(T) \ge Q(T_1) (1+ \varepsilon)^{T- T_1}.
\end{equation}
Let us note that $Q(T_1) \ge p_0$ and $Q(T) \le  1/p_1$.Therefore, one has
\begin{equation} \label{QQ1}
\frac{1}{p_1  p_0}\ge   (1+ \varepsilon)^{T- T_1}.
\end{equation}
This inequality leads to a contradiction for $T=T_1+T_c$ and $T_c$ satisfying (\ref{zzz}) that finishes the proof.  
\vspace{0.2cm}

Let us make some comments.  The assertion is not valid if the set $Diff(s, s^{`})$ 
belongs to two different regulation sets $K_i, K_j$. This effect is connected with a pleiotropy in the gene regulation.   However,  if assumption {\bf M2} holds then the pleiotropy probability is small for large genome lengths $N$. 

Moreover,  Theorem II can be extended on the case of finite but large populations when the genetic drift effect is small and under the assumption that the mutation probability also is small.  
We obtain  the following assertion:

{\bf Evolution Recognition Theorem   III} \label{T3a}

{\em Consider the population dynamics defined by model {\bf 1}-{\bf 3} in subsection \ref{popdyn}. Assume  conditions {\bf A}, {\bf B} and {\bf M2} hold,  and assumptions  
(\ref{zzz1}), (\ref{Diff}), (\ref{kappa}), (\ref{zincr}) of the previous theorem are satisfied. Suppose $T_c$   satisfies the inequality
\begin{equation} \label{zzz1a}
T_c > \frac{-  \log(p_0 p_1)}{1 +\varepsilon/2},  \quad \varepsilon= \frac{\kappa \delta}{S_b}. 
\end{equation}
We suppose that the population size satisfies
\begin{equation} \label{Npmin}
N_{\mathrm{pop}}(t)  \ge   N_{\mathrm{popmin}}   \quad \forall t \in [T_1, T_1+T_c]. 
\end{equation}
and
\begin{equation} \label{zzz2a}
T_c \ll  \log(N_{\mathrm{popmin}}).
\end{equation}

Let the mutation frequncy $p_{\mathrm{mut}}$ be small enough 
the population abundance
$N_{\mathrm{popmin}}$ be sufficiently large so that
\begin{equation} \label{pmutb}
 \frac{1 +\varepsilon)(1 -\rho_0) - 2p_{\mathrm{mut}} S_B /F(s')}{1+ \rho_0 + 2N_{\mathrm{popmax}} p_{\mathrm{mut}}(p_1 N_{\mathrm{popmin}})^{-1}}
 > \varepsilon/2,  \quad \rho_0=N_{\mathrm{popmin}}^{-1/4}. 
 \end{equation}
 and
 \begin{equation} \label{pmutb1}
 p_m=N_{\mathrm{popmax}} p_{\mathrm{mut}} \ll 1.
 \end{equation}
Then if $j \in I_{-}$ the inequality 
\begin{equation} \label{ZZ}
  p_1  < p_0^{-1} (1 + \varepsilon/2)^{T_c}
\end{equation}
is satisfied with the probability 
\begin{equation}  \label{beviable}
Pr_{v} > (1 -  4 P_{*} p_m)^{T_c}  
\end{equation}
where
\begin{equation} \label{Prob1}
P_{*}= \exp(-  0.5  F(s')S_B^{-1} N_{popmin}^{1/2}).
\end{equation}

}

{  %We proceed the proof under additional technical assumption that $F(z(s')) < 2$.   
\vspace{0.2cm}

{\em Statistical interpretation of the Theorem}

This theorem shows that evolution can make a statistical test checking the hypothesis $H_{-}$ that
$j \in I_{-}$ against the hypothesis $H_{+}$ that $j \in I_{+}$. Our arguments repeat classical reasoning of mathematical statistics. Namely, suppose  $H_{-}$ is  true. Let $V$ be the event that  frequency of the genotype is larger than $p_1$  is viable
within a sufficiently large checking time $T_c$.  According to   estimate  (\ref{beviable}) 
the probability of this event  $V$ is so small that it is almost unbelievable.  Therefore, the hypothesis
$H_{-}$ should be rejected.

\bf Proof}.

\vspace{0.2cm}
{\em Ideas beyond proof}.  Actually the main idea is the same that in the previous theorem: compare
the abundances of mutants with the genotypes  $s'$ and individuals with the genotype $s$ in the population. However, the proof includes a number of  technical details connected with estimates of mutation effects and fluctuations of the abundances of different genotypes. 
\vspace{0.2cm}

Proof can be found in Appendix 1.

%\vspace{0.2cm}

\medskip\noindent
We will refer  $T_c$ as {\em checking time}.  The sense of this terminology becomes clear  in the next subsection.

\subsection{Learning by gene networks for random fitness landscapes}

In this section we describe  how the genetic regulation network can
perform the landscape fitness learning.

 First we consider  model defined by (\ref{cellj}) and (\ref{envW}).  
Assume that $f_j(s(t+1), h_j) > f_j(s(t), h_j)$ for some $i$ and a time moment $t$ as a result of a mutation, 
but nonetheless the organims is viable up to the moment $t + T_c$. According to Theorems II and III this fact
indicates that, with a probability close to $1$,  the corresponding coefficient $b_j  >0$. In fact, if the mutation probability is small enough and the population has a sufficiently large size, then $j \in I_+$, i.e., $b_j >0$. 

Therefore,  only under condition $f_j=1$    the fitness attains the maximal   value. 

The key idea is as follows. In order  to obtain $f_j=1$ evolution has two diffrent ways. The first way is 
to make mutations in the genes $s_i$ involved in expression 
$S_1=\sum_{i} w_{ji} s_i - h_j$.  If there are a number of such genes it is a longtime way. Another way is just to vary the threshold  $h_j$.  
 This regulation of $h_j$ can be performed by the feedback mechanism (\ref{barhy}). 
We take positive values $g_{ik}$ and a large $r_F$, and adjust parameters in (\ref{yH}) in such a way  that
$y_k \approx 1$. 
This regulation mechanism is not gradual and it may be faster than mutations in all genes involved in the
expression of the corresponding phenotypical trait.  
Such regulation directs (canalizes) evolution to a ``correct'' way sharply reducing the number of mutations.   That evolution mechanism organized by the gene regulation via thresholds we will refer as
{\em  canalized evolution}.  Consider an example. 
 
Let a trait be regulated by, say, $10$ genes, involved in a threshold mechanism of  relation defined by
 (\ref{cellj}). Suppose  the corresponding $w_{ji}=1$ and the threshold $h_j=5$. Moreover, let at initial moment all genes involved in regulation are not expressed $s_i=0$.
Then usual random mutation and selection evolution leading to maximal expression needs at least $5$ mutations and 
the corresponding time is $T_{D} \approx p_{\mathrm{mut}}^{-5}$
(the index $D$ in honor of Darwin).

For the canalized evolution the first successful mutation is only a test.  If $f_j$  increases as a result of a mutation and the mutant organism is viable within 
a large  checking period $T_c$,
the gene regulation maximally decreases the corresponding threshold $h_j$. 
Thus the canalized evolution leads to the maximal expression within the time $T_{W}=p_{\mathrm{mut}}^{-1} + T_c$ (the index $W$ in honor of Waddington). 
According to estimates of Theorems II and III,   $p_{\mathrm{mut}}  \ll  \log(N_{\mathrm{popmax}})$ and then one has
$T_D  \gg  T_W$.

Consider now the general case where the fitness is defined by (\ref{Fi}) and (\ref{envW}).   First we  use the methods of statistical physics to estimate
a learning error.  We are going to show that learning can sharply reduce the number of mutations 
to attain the maximal fitness  even when  the coefficients $b_j$ in the fitness potential $W$ are random. More precisely,
we assume that $b_{\bf j}$ are mutually independent random coefficients distributed according to
a probabilistic measure $\d\xi_{ j}(x)$ on the space ${\bf R}$.  Let $N_b=m$ be the number of coefficients $b_{ i}$.  
Then we have the product measure $\d\mu=\prod_{ j}  \d\xi(b_{ j})$
on  ${\bf R}^{N_b}$. According to the Evolution Recognition Theorem,  one can expect thus that
 $F(s(t_0)) \le F(s(t_1))$ for the most of population members and $t_1 - t_0  \gg  T_c$.  
 Here we can also apply  Fisher's theorem, which asserts  that the averaged fitness increases in time
(see the end of subsection  \ref{popdyn}).

 Let us set  $s^i=s(t_i), \ i=0,1$. Let us denote by $f$ the vector with components $f_i$:
$$
f(s, h)=(f_1(s, h_1), \ldots .,  f_{N_b}(s, h_{N_b})),
$$
which can be interpreted as a ''phenotype", 
and let $f^i=f(s^i, h)$.
 Let  $P(f^0, f^1)$ be the probability of  the event $Y_0$  that $F(s^0, b) \le F(s^1, b)$.  Then
 \begin{equation} \label{P1}
 P(f^0, f^1)=Pr\{Y_0 \}=\int_{{\bf R}^{N_b}}  
 \chi_{W(f^0) > W(f^1)}(b) 
 \d\mu(b), 
\end{equation}
where $\chi_{\mathcal C}$ is a characteristic function of the set defined by some condition $\mathcal C$ and 
$$
W(f)= \sum_{i=1}^{N_b} b_i  f_i.
$$
For each pair of vectors $f^0, f^1$ we introduce the vector $\psi(f^0, f^1)$ with $m=N_b$ components by the relation 
\begin{equation} \label{psidef}
\psi_{ i}(f^0, f^1)=f_{ i}^0  - f_ i^1.
\end{equation}
Note  that $P(f^0, f^1)$ is the volume of the half-space defined  by the hyperplane, which goes
through the origin ${\bf 0}$ and has the normal vector parallel to $\psi(f^0, f^1)$.

Consider the event $Y_1$ that
for $f=f^*$ we have   $W(f^*, b) > W(f^1, b)$ and
the conditional probability 
 \begin{equation}
 \mathrm{Pr}(Y_1|Y_0)=\mathrm{Pr}(Y_0 Y_1)/\mathrm{Pr}(Y_0)=I(f^*, f^0, f^1). 
\end{equation}
Note that $\mathrm{Pr}(Y_0 Y_1)$ equals the volume of the intersection $\mathcal S$ of the two
half-spaces bounded by the two hyperplanes going through the origin and orthogonal 
to $\psi(s^0, s^1)$ and $\psi(s^1, s^*)$, respectively.  The landscape learning risk can be defined by
$$
R(f^*, f^0, f^1)=1 - \mathrm{Pr}(Y_1|Y_0)=1 -  I(f^*, f^0, f^1). 
$$
Let the measure $d\xi_{ j}$ be  $d\xi_{j} =\rho_{\bf j}(x) dx$, and $\rho_{ j} \in {\bf N}(0, \lambda_{j})$ be a normal density with the variation $\lambda_{\bf j}^{-2}/2  >0$.
We introduce important auxiliary quantities
\begin{equation} \label{a00}
a_{00}=|| \psi(f^0, f^1)||^2,  \quad   a_{11}=|| \psi(f^1, f^*)||^2,
\end{equation}
\begin{equation} \label{a01}
a_{01}=\langle \psi(f^0, f^1), \psi(f^1, f^*) \rangle
\end{equation}
where $|| \   ||$ and $\langle, \rangle$ denote the norm and the scalar product in the $N_b$ dimensional Euclidian space ${\bf R}^{N_b}$ defined by
$$
    \langle \psi , \phi \rangle =\sum_{ i=1^{N_b}}  \psi_{ i} \phi _{ i} \lambda_{i}^{2}, \quad ||\psi||^2=\langle \psi, \psi \rangle.
$$

{\bf Theorem IV} (on Learning  Risk)  \label{T4}
{\em   
Let $d\mu=\prod_{\bf j}  d\xi_{j}$, where $d\xi_{ j}$ are defined above by normal densities. 
Then  the risk $R=R(f^*, f^0, f^1)$ can be computed by
\begin{equation}  \label{Iz}
R= \pi^{-1} \int_0^{+\infty}  \d y \exp(-y^2/2)\int_{-\infty}^{-\omega y} \exp(-v^2 /2) \d v,
\end{equation}
where
\begin{equation}  \label{Iz}
\omega= \frac{a_{01} }{a_{00} r}, \quad r= \sqrt{a_{00} a_{11} - a_{01}^2}.
\end{equation}

}
{Proof}  uses special methods of statistical physics, see Appendix 2.

{\em Comment.} Numerical computations show that $R(f^*, f^0, f^1) \approx 0.5 \omega^{-1}$.  This result has a transparent geometrical interpretation. 
For small $r$ the quantity $\omega  \approx r^{-1}$ thus   $R(f^*, f^0, f^1) \approx 0.5 r$.  If $\psi(f^0, f^1)=\psi(f^1, f^*) +\tilde \psi$, where
$||\tilde \psi||  \ll  ||\psi(f^0, f^1)||$ then 
\begin{equation} \label{angr}
r \approx   || \psi(f^0, f^1)|| \  || \tilde \psi|| \sin \theta,   
\end{equation}
$$
\cos \theta=\frac{  \langle \tilde \psi,  \psi(f^0, f^1) \rangle}{   || \psi(f^0, f^1)|| \ || \tilde \psi||}
$$
and $r$ involves two factors.  The first factor $f_1=|| \psi(f^0, f^1)|| || \tilde \psi||$ is connected with the ruggedness  of the fitness landscape.
 The second factor $\sin \theta$ is proportional to the angle between the hyperplanes mentioned above. 
At each evolution step, the  learning risk is proportional to the volume of a multidimensional cone restricted by the two hyperplanes.

According to    Theorem IV,
our regulatory mechanism should find a point $s_*$ such that
$I(f^*, f^0, f^1)$ is maximal (and $R(f_*, f^0, f^1)$ is minimal).  Note that
$I(f^*, f^0, f^1)=1$ for $r=0$, i.e., when the vectors $\psi(f^0, f^1)$ and $\psi(f^1, f^*)$ are parallel. These vectors are almost parallel for small regulation parameters $r_F$.
Thus we obtain the following 

{\bf  Regulation rule.} 

{\em Given $s^0$, $s^1$, the regulatory mechanism should find a value  $h^*$
such that the angle $\theta(s^*)$ between vectors $\psi(f^0, f^1)$ and $\psi(f^1, f^*)$ is minimal, where
$$
f^i=f(s^i,  h),  \quad   f^*=f(s^1, h^*).
$$
}

In the next subsection we consider a possible biological  realization of this regulation rule. 

\subsubsection{Gene regulation scheme for learning}

In the general case we  apply the time-recurrent regulatory scheme defined by  (\ref{barhy}) and (\ref{yH}) with a small 
$|r_F|$. Let us introduce 
$$
S(\alpha, h^*)=|| \alpha(f(s^1, h) -  f(s^0, h)) - f(s^1, h^*)+ f(s^0, h)|| .
$$
According to the regulation rule  and definition (\ref{psidef}) of $\psi$,  
$h^*$ should minimize $S(\alpha, h^*)$     
for some $\alpha >0$.  We assume that $h^*=h(s^1)$ and 
$h=h(s^0)$, where $h(s)$ is defined by 
(\ref{barhy}) and (\ref{yH}).  Our problem reduces then to a correct definition of parameters $a, \tilde h, \tau_r$ and $g$  in 
the gene regulation scheme.  This problem can be resolved in the case (\ref{cellj}) under assumption that gene redundancy is large, i.e., $\beta \gg 1$ and each trait $f_j$ is controlled by many genes.
As it will be shown below,  the solution admits an interesting biological interpretation.   
 
We set $h_j^*=h_j +\Delta h_j$.  Assuming that $\Delta h_j$
and $\alpha$ are small, we obtain

\begin{equation} \label{reg}
 \alpha(f_j(s^1, h_j) -  f_j(s^0, h_j)) \approx \frac{\partial f_j(s^1, h_j)}{\partial h_j} \Delta h_j.
\end{equation}

In the case (\ref{cellj}) and for a large redundancy $\beta$
the last relation reduces to   simpler relations, namely,
\begin{equation} \label{g1}
\alpha \sum_{i=1}^N w_{ji}(s_i^1 -s_i^0) \approx  \Delta h_j.  
\end{equation}
This relation will be fulfilled  if we choose parameters $a, g, \tilde h$ as follows  (there exist
many other choices). 
Let us consider the case  $\sigma_R= a_R x$, i.e., a linear feedback and $\tau_r=T_c$.  We set
$\tilde h_j=0$ and $g_{ik}(\tau)=0$ for all $\tau <  T_c$. Let us set $b_{il}=-\sum_{k=1}^n 
g_{ik} a_{kl}$, where  $i=1,\ldots , m=N_b$ and $l=1,\ldots , N$. Then relation (\ref{g1}) can be rewritten as follows:

\begin{equation} \label{g2}
\alpha \sum_{i=1}^N w_{ji}(s_i^1 -s_i^0) = r_F  \sum_{i=1}^N b_{ji}(s_i^1 -s_i^0), 
\end{equation}
for all $j=1,\ldots , m$.

Let consider matrix $B$ with entries $b_{ij}$ and the corresponding the sign matrix $S(B)$
with entries $s_{ij}(B)=sign(b_{ij})$.  The matrix $W$ introduced above in the end of subsection  \ref{trait}
determines phenotype control by the genes while $B$ defines a gene regulation scheme for control of the thresholds $h_j$, i.e., thresholds that deterimine the rates of gene expression. 

 The next claim states an important connection  between $S(B)$ and $S(W)$.
\vspace{0.2cm}

{\bf Assertion}. {\em  Let $S(W)=S(B)$. Then, if all mutations are point ones,  they are 
seldom enough, and assumption {\bf M2} is satisfied,
then relation (\ref{g2}) holds  for some $\alpha$ with probability close to $1$}.

\vspace{0.2cm}

{\bf Proof}.
In fact, if mutations are sufficiently seldom, then Hamming distance between $s^1$ and $s^0$ is $1$,
i.e., vector $\tilde s=s^1 -s^0$ contains a single $1$ ( or $-1$) and all the rest components 
of $\tilde s$ equal $0$.  Let $\tilde s_i=1$ for $i=i_*$. Note that
the most of the coefficients $w_{ij}$ in the sums  $\sum_{i=1}^N w_{ji}(s_i^1 -s_i^0)$ 
are equal zero.  Due  to {\bf M2} we should satisfy (\ref{g2}) only for a single $j$, say, $j=j_*$
(it is valid with a probability close to $1$). This index
corresponds to the phenotypic trait affected by a mutation in $s_{i_*}$.
Then, if $S(W)=S(B)$,  we can set   $\alpha =r_F^{-1}  w_{j_* i_*}^{-1} b_{j_*  i_*}$. 
\vspace{0.2cm}

{\em Remark}: this proof show that $\alpha(t)$ could be  different for different evoloution moments $t$.

\subsubsection{Biological corollaries: Phenotype control, gene regulation and evolution} 

The last assertion has interesting biological consequences. If mutations are seldom, and pleiotropy in phenotype gene regulation is weak (assumption {\bf M2}) then the fitness landscape learning is possible under the condition that  $S(W)=S(B)$.  If we translate the last fact  on a biological language, this means  the following.  Suppose that  $w_{ji} >0$, i.e,
activation of $i$-th gene reinforces expression of $j$ -th phenotypic trait.  Then also 
$b_{ji} >0$ , i.e, activation of    $i$-th gene reinforces expression of $j$ -th threshold $h_j$. Similarly, if $w_{ji} < 0$, i.e,
activation of $i$-th gene represses expression of $j$ -th phenotypic trait.  Then also 
$b_{ji} >0$ , i.e, activation of    $i$-th gene represses expression of $j$ -th threshold $h_j$.

 Remind that the matrix $W$ defines the phenotype control whereas $B$ defines the gene regulation of this control, i.e.,  evolution can perform the fitness landscape learning {\em without too refined tuning of gene regulation, with a rough tuning only} since
 the relation $S(B)=S(W)$ implies only a rough correspondance between $W$ and $B$. The matrix $W$ defines
the morphogenesis via gene expression and $B$ is connected with gene regulation. One can think that the matrix $B$ is a result of evolution. 

So,  the rough correspondence between $W$ and $B$ via $S(W)=S(B)$
means that if organisms are capable to the fitness
landscape learning, then   the phenotype  organization via genes  repeats, in general times,
 the gene regulation developed by evolution.  One can say that   the morphogenesis  recapitulates evolution,  an idea     intensively discussed in 19 and 20 centuries.

\subsubsection{Estimate of learning risk for many regulation steps}

By relation (\ref{angr}) 
and some computaitions we can find that  the learning risk $R$ at the evolutionstep $t$  has the order $ r_F^2$, where $r_F$ is the parameter that determines the magnitude of the feedback in the gene regulation scheme, see (\ref{barhy}).
Thus the learning risk is small for a weak regualtion. At each regulation step, the probability that the learning rule does not make an error, is 
$\approx 1- c_1 r_F^2$.   The number of need steps can be estimated as  $const r_F^{-1}$.  Therefore the total risk is $R_{tot} \approx (1 -c_1 r_F^2)^{c_2 r_F^{-1}}$, where
$c_k>0$ are constants as $r_F \to 0$  and $R_{tot}$  tends to $0$ as $r_F \to 0$.
 We conclude that for the fitness functions with random $b_i$ the canalized evolution consisting of many trials and errors and  going in  a gradual manner leads to a small total risk.

\subsection{Acceleration} \label{Hebb}

\subsubsection{Reduction of mutation number   for a monotone fitness}

Suppose that we have a  mutation  such that the expression of $f_j$ increases.  Due to theorems II, III with a probability close to 1 we have then that for the corresponding gene  $i \in I_{+}$, i.e., this gene lies in the positive set. 
This means
that a successive growth in $f_i$  expression increases viability of the organism. Let us compute the number of mutations $\mu(h_i, L)$, 
which are necessary to express  of $f_i$ at the level $L$: $f_i=L$.  This number can be computed for  model defined by (\ref{cellj}) and (\ref{envW}).

Assume for simplicity that all $w_{ij}$ equal $w>0$ or $0$ and we have at least $k_j$ non-zero $w_{ij}$ where $k_j > h_j$.  Moreover, we suppose that all $s_i$ with $i \in K_j$ 
equal $0$ at the initial time moment. Let  $S_L= \sigma^{-1}(L)$, where $\sigma^{-1}$ is a function inverse to $\sigma$. We use the ceiling  function $[x]$, which  maps a real number
$x$ to  the smallest following integer.  

Then the need mutation number $\mu_i$ is 
\begin{equation} \label{mui}
\mu_i= [(S_L +  h_i)/w],
\end{equation}
i.e., it is a  non-negative integer closest to $(S_L +h_i)/w$ and greater than $(S_L +h_i)/w$.  
If the sigmoidal function $\sigma$ is close enough to the  step function, then $\mu_i$ is a non-negative integer closest to $h_i/w$ and larger than $h_i/w$. 

A decrease of $\mu_i$ sharply increases chances to survive.  According to   (\ref{mui}) to decrease $\mu_i$ we should decrease $h_i$.

\subsubsection{Acceleration of evolution by gene regulation }

An estimate of evolution acceleration follows from the arguments of the previous subsection. We again consider model (\ref{cellj}), (\ref{envW}).  Let us compare the two cases: when $r_F=0$, i.e, we have no gene regulation, and $r_F>0$.   Suppose that $y_k$ are close to $1$. Denote by $\mu(L)$  the total number of mutations need to attain expression level $L$
for all  $f_i$.  
One obtains
\begin{equation} \label{muiT2}
\mu(L) = \sum_{i=1}^m [(S_L + \bar h_i)/w].
\end{equation}
Therefore, 
the evolution times  is given by relation
 \begin{equation} \label{timeac}
T_{D} \approx \mathrm{const} p_{\mathrm{mut}}^{-\mu(L)},  
\end{equation}
where $T_{D}$ is the time necessary to reach a need expression level of all phenotypic traits $f_i$.

For $r_F >0$ we apply the learning algorithm described above and note that we need $m$ mutations only. Let us denote 
 $T_{W}$   the averaged time to reach that expression  by a "canalized" landscape learning process.
Then
\begin{equation} \label{timeac2}
 T_{W} \approx p_{\mathrm{mut}}^{-m}  +  \mathrm{const} T_c,  
\end{equation}

The time acceleration is $T_{D} -T_{W}$  if that quantity is positive.  Note that $T_c$ is proportional to
a logarithm of $ 
p_{\mathrm{mut}}$, so one can expect $T_D \gg T_W$.

\subsection{Acceleration of evolution for general case}

%Therefore, the total  mutation number $\mu$ is a linear function of the feedback parameter $r_F$. The probability $p_L$ to attain the expression level $L$ 
%is then an exponential function of the feedback parameter $r_F$ because $p_L=p_{\mathrm{mut}}^{\mu}$. 
For general case an estimate of the evolution acceleration can be done under the following assumptions.

{\bf Assumption C1}. {\em  For the expression  level $L$ of trait $f_j$ for all $s$ there exists  a threshold value $h_j^*(s)$ such that $f_j(s, h_j^*) > L$}.

{\bf Assumption C2} {\em  For $r_F=0$, $h_j=\bar h_j$  and a random genotype $s$ at least $\mu_j$ mutations in $s $ are necessary to reach expression level $L$}.

Let $\mu(L)=\sum_{i=1}^m \mu_i$. 
Then 
\begin{equation} \label{timeac1}
T_{D} \approx \mathrm{const} p_{\mathrm{mut}}^{-\mu(L)},  
\end{equation}
\begin{equation} \label{timeac2}
  T_{W} \approx p_{\mathrm{mut}}^{-m}  +  \mathrm{const} \max_j  (h_j^* - \bar h_j)r_F^{-1} T_c.
\end{equation}

To conclude this subsection let us note that there is an interesting biological uncertainty principle.  Namely, to decrease the learning risk, evolution should use more
test rounds and a smaller regulation parameter $r_F$, however, for smaller $r_F$ we obtain greater evolution times $T_W$.  So, the canalized evolution increases
the risk to come in an evolution dead end.

\subsection{Phenotype canalization}

We consider the case when the $f_j$ expression is defined by (\ref{cellj}).
To understand how the feedback parameter $r_F$ affects the robustness of the phenotype $f$ with respect to mutations, let us observe that for typical sigmoidal functions
$\sigma(S)$ the derivative  $\sigma^{'}=d\sigma(S)/dS$ is a decreasing function of $S$ (this property holds for $\sigma=1+ \exp(-x)$ and other examples).   In the general case,
$d\sigma(S)/dS$ is a decreasing function of $S$ for sufficiently large $S$ (it follows from the limit condition $\sigma(+\infty)=1$). The robustness is defined by  the quantities
\begin{equation} \label{celljr}
R_j = \sigma^{'}(\sum_{i=1}^N w_{ji}{s_i} -  h_j(r_F)).
\end{equation}
For smaller $R_j$  the robustness and the canalization effect  in $f_j$-expression are  stronger.
Relations (\ref{cellj}) and  (\ref{barhy}) show that the robustness increases in $r_F$, i.e., the feedback  based on (\ref{yH}) and (\ref{barhy})  accelerates evolution  and stabilize patterns.
For general $f$ the same conclusion can be obtained  if the second derivative  $f^{''}$ of $f_j$ with respect to $h_j$ is negative for large $h_j$.  

This effect of the robustness increase  has an interesting consequence if we consider a varying environment and
the fitness depends on a time $t$ slowly. For example, we can assume that coefficients $b_j$ in (\ref{cellj}) depends on a slow time 
$\tau$. Suppose that at a time interval $[T_1, T_c]$ the fitness is larger if a trait (morphogene)
$f_k$ is expressed completely, $f_k \approx 1$ but for $t > T_2 > T_c$ the situation is opposite, i.e.,  the fitness is greater for $f_k \approx 0$.  It is clear that  inverse mutations in $f_k$
which decrease of $f_k$ expression, increase  survival chances.  However, a strong canalization diminishes the probability of such inverse mutations.  So, canalization can lead to an evolution dead end, when too specialized species extinct as a result of an environment change.

\subsection{Numerical simulations}

\subsubsection{Simulations for simplest model}

In simulations, to investigate  a large set of fitness  models, we use the variables $z_j$ defined by (\ref{cellj}). 

 First numerical model (\ref{envW}) 
 $$
 F(s)=C_M \exp(\lambda W), \quad W=\sum_{j=1}^m b_j z_j(s)),
 $$
 with  $b_i >0$ is tested.  To simplify calculations we make only point mutations, each gene can be flipped
 with probability $p_{\mathrm{mut}}$.

The parameters were as follows. The number of genes $N=40$,  $C_M \approx 1$, the population size
$N_{\mathrm{pop}}=50$ and  the maximal population size $N_{\mathrm{popmax}}=60$. We have made $100$ evolution steps. The parameter $\lambda$ equals $0.2$, the mutation probability $p_{\mathrm{mut}}=0.05$, the feedback parameter
 $ r_F \in (0,1)$.
Note that  this population is not large, and formally our Theorems II, III are not applicable.  The landscape learning for small populations may be particularly interesting for the bottleneck problem. 
A population bottleneck (or genetic bottleneck) is a sharp reduction in the size of a population due to environmental events (such as earthquakes, floods, fires, disease, or droughts) or human activities (such as genocide). Such events can reduce the variation in the gene pool of a population that increases genetic drift affect. Notel that equations  (\ref{popdyn1}), (\ref{popdyn2})  do not take into account this effect. 

In numerical simulations we have used the following simplified variant of gene regulation.
We  set $a_{ij}=a_0 >0$. If  for an evolution step  $t$ one has $z_j(t) > z_c$ we set $h_j=\bar h_j - r_F$. If $z_j(t) < 1-z_c$, we set $h_j=\bar h_j + r_F$.

Here $r_F$ and $z_c$ are positive parameters,  in simulations the values $r_F=1$ and $z_c=0.7$ are taken.  The fitness was defined by relation (\ref{Fi}) with $L=9$ and
random positive $b_i \in (0,1)$.

To measure canalization and robustness we introduce the following numerical characteristics $S$. 
This quantity measures sensitivity  of the fitness with respect to random mutations. Let $s=(s_1, \ldots  ,s_N)$ be a genotype and $s'$ be another genotype  which differs from $s$ by a flip at randomly chosen position. We set $S=\Delta F/F(s)$, where $\Delta F=F(s)-F(s')$. 

The numerical results are well consistent with analytical conclusions and can be illustrated by the following plots.

\begin{figure}[h!]
\vskip-0.5truecm
\includegraphics[width=80mm]{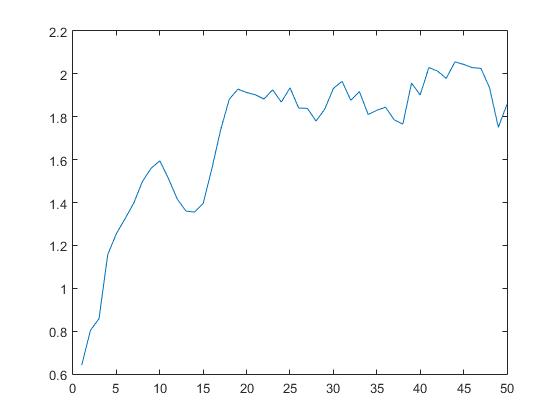}
\caption{\small This image shows the mean fitness potential as a 
function of evolution time step. We observe an increase of the fitness}\label{Fig1}
\end{figure}

\begin{figure}[h!]
\vskip-0.5truecm
\includegraphics[width=80mm]{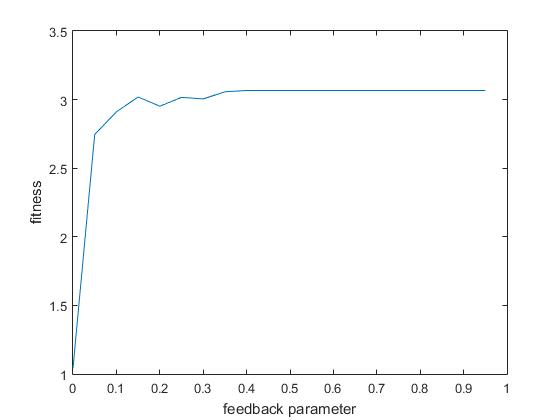}
\caption{\small This plot shows the mean population fitness at the final evolution step as a function of  the feedback parameter $r_F$. Note
that for small $r_F$ the fitness does not reach maximum.}\label{Fig2}
\end{figure}

\begin{figure}[h!]
\vskip-0.5truecm
\includegraphics[width=80mm]{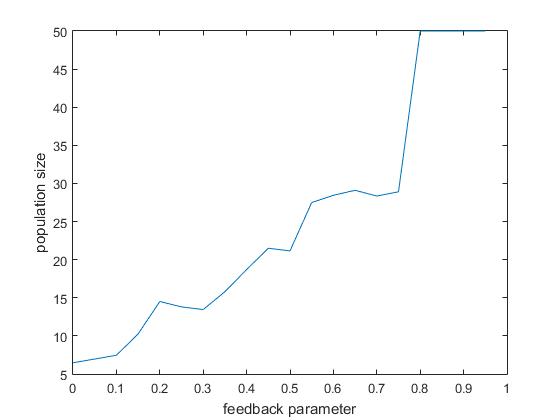}
\caption{\small This plot shows the population abundance at the final evolution step as a function of  the feedback parameter $r_F$. For small $r_F$ only a small part of organisms survive while for $r_F \approx 1$ all are viable. }\label{Fig3}
\end{figure}

\begin{figure}[h!]
\vskip-0.5truecm
\includegraphics[width=80mm]{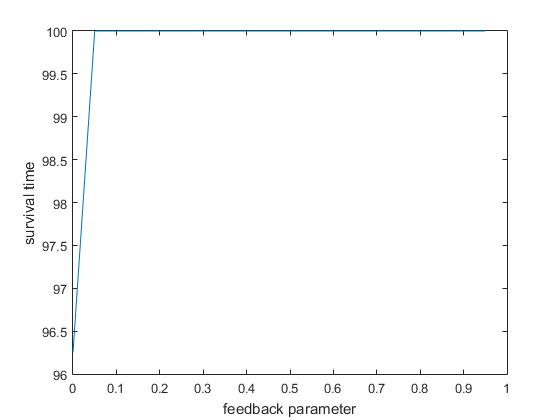}
\caption{\small This plot shows the mean survival time as a function of  the feedback parameter $r_F$. For small $r_F$ many populations extinct.}\label{Fig4}
\end{figure}

\begin{figure}[h!]
\vskip-0.5truecm
\includegraphics[width=80mm]{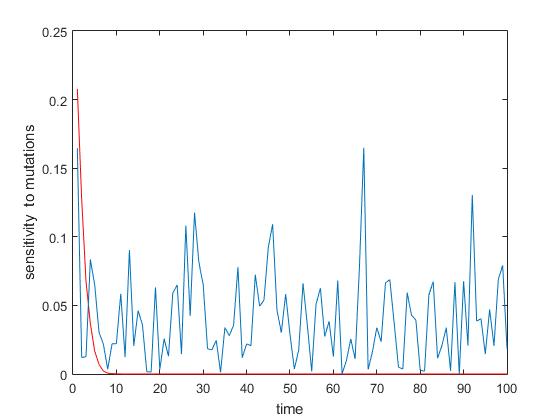}
\caption{\small This plot  shows two sensitivities as functions of time. The red curve shows  sensitivity when the regulatory gene network canalizes evolution  and $r_F=1$. The blue curve shows  sensitivity when $r_F=0$ and the regulatory gene network is not involved in evolution.  We see that the effect of gene regulation is very strong.}\label{Fig5}
\end{figure}

\subsubsection{Simulations for different fitness functions}

In the next series of simulations we consider fitness functions defined by $F=C_M \exp(\lambda W)$ with different potentials $W$. 
The main idea is to check a capacity of gene regulation to perform the fitness landscape learning 
even for very sophisticated fitness functions.

In  all the cases
(except for  the Rosenberg function, see below) we use the simplest regulator described above and $C_M=5,  \lambda=1$, $N_{\mathrm{popmax}}=100$ and $p_{\mathrm{mut}}=0.05$. In the canalized case the regulation parameter $r_F=1$,
if canalization (regulation)  is absent,  we set $r_F=0$.  

{\bf a} )  Let  $W$ be defined by (\ref{envW})  with $b_i =1$, $L=4$,  and $\mu=5$, $m_I=2$. 
  The results are shown on Fig. \ref{Fig1}.

\begin{figure}[h!] \label{Fit0}
\vskip-0.5truecm
\includegraphics[width=80mm]{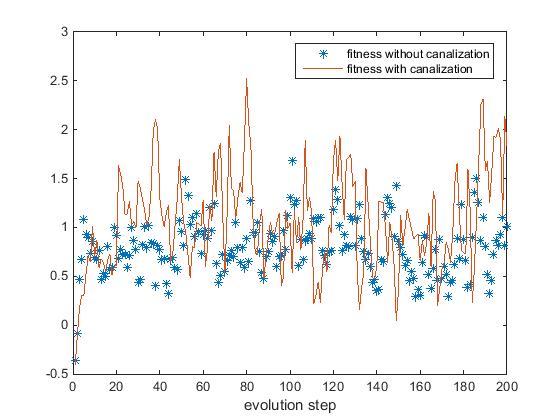}
\caption{\small The case {\bf a}). This plot shows that regulation with canalization increases the fitness.
The fitness is shown in a logarithmic scale, i.e., the vertical axis shows the fitness potential $W$.}
\end{figure}

{\bf b} )  In this case we set $W= \sum_{i=1}^2 (z_i-1)^2 + \sum_{i=3}^6  z_i^2$.  This function has a single  maximum on the hypercube $[0,1]^6$ 
at $(0,0, 1,1,1,1)$.

\begin{figure}[h!]   \label{Fit1}
\vskip-0.5truecm
\includegraphics[width=80mm]{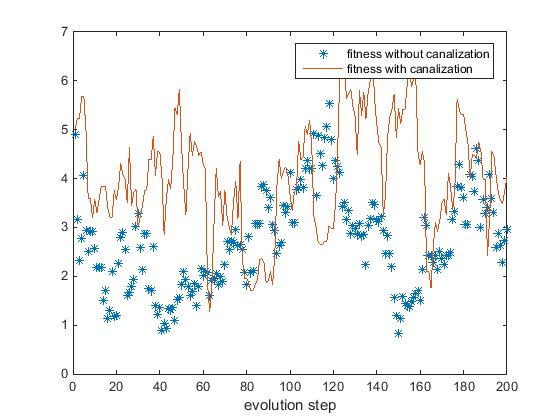}
\caption{\small  The case {\bf b}). This plot shows how regulation with canalization increases the quadratic fitness with a single local maximum. The fitness is shown in a logarithmic scale, i.e., the vertical axis shows the fitness potential $W$.}
\end{figure}

{\bf c} )  In this case we set $W= \sum_{i=1}^6 b_i (z_i-1)^2 + 0.8b_i z_i^2$.  This function has   local maxima on the hypercube $[0,1]^6$ 
at the all hypercube vertices. 

\begin{figure}[h!]   \label{Fit2}
\vskip-0.5truecm
\includegraphics[width=80mm]{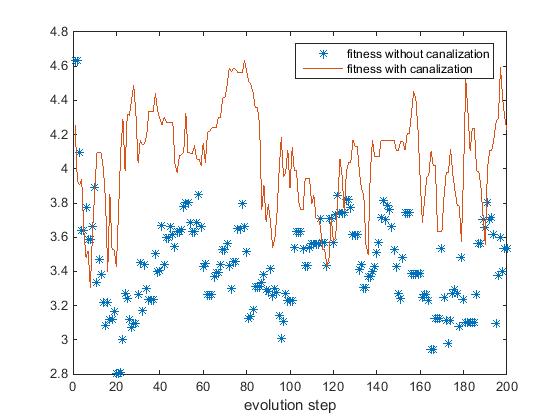}
\caption{\small The case {\bf c}). The fitness is shown in a logarithmic scale, i.e., the vertical axis shows the fitness potential $W$.}
\end{figure}

{\bf d})     The  quadratic fitness potential   $W=\sum_{j=1}^6 b_j*(z_{2j-1} + z_{2j} - 2z_{2j-1}z_{2j})$, where there exist a number of local maxima.   
\begin{figure}[h!]
\vskip-0.5truecm
\includegraphics[width=80mm]{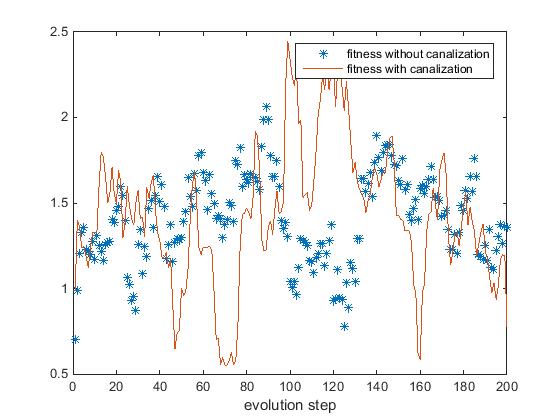}
\caption{\small The case {\bf d})}\label{Fit3}
\end{figure}

{\bf e}) Let us consider  a special fitness,  defined by the Rosenbrock function, which serves as a test for different optimization algorithms.
The Rosenbrock function has the form
$$
W=\sum_{j=1}^m (1- z_ j)^2 + a(z_{j+1} -z_j^2)^2,
 $$
where $a>0$ is   a large parameter.  It is also known as Rosenbrock's valley because
the global minimum is inside a long, narrow, parabolic shaped valley. To find the valley, it is easy but to attain the global minimum, it  is hard.
Indeed, numerical simulations for $m=5$ and  $a=10$ show that the fitness does attain the global maxima within  a long evolution  ($> 200$ steps),
in  the both cases: without any regulation and with one described above in the beginning of this subsection.  However, if we apply more sophisticated regulation defined by 
an analogue of the Hebb rule,   we obtain a fast convergence to global maximum at $z=(1,1,1,1,1,1)$, see Fig. \ref{Fig5}.
  So, we can go through valley by a gene learning organized by a Hebb's like  principle.  How evolution can cross a fitness valley 
 is an important question in genetics and it was considered by \citep{Weiss}.

\begin{figure}[h!]
\vskip-0.5truecm
\includegraphics[width=80mm]{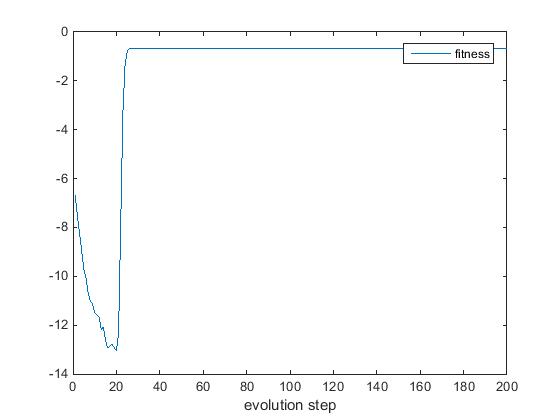}
\caption{\small The case {\bf e}. The fitness potential $W$  approaches to  the global maximum for a hard test fitness function if evolution learning is organized in a special way. The fitness is shown in a logarithmic scale, i.e., the vertical axis shows the fitness potential $W$. }\label{FitRH}
\end{figure}

\section{ Conclusions and Discussion}

The famous computer scientist \citet{Val2} wrote that evolutionary biology cannot explain the rate at which evolution occurs: 
``The evidence for Darwin's general schema for evolution being essentially correct is convincing to the great majority of biologists. 
This author has been to enough natural history museums to be convinced himself. 
All this, however, does not mean the current theory of evolution is adequately explanatory. 
At present the theory of evolution can offer no account of the rate at which evolution progresses 
to develop complex mechanisms or to maintain them in changing environments.''

The first attempt to explain the fast evolution rate was made by %Waddington 
\citet{Wad, Wad1}, who gave an idea of canalized evolution, however, without
suggesting any mathematical models. Recently, however, some ideas and models 
were proposed by \citep{Valiant2006,Valiant, SatE, FV}, see also \citep{Watson} for a review,  in order 
to explain a possibility of adaptation with formation of many phenotypic traits.  
These models exploit basic Computer Science Theory approaches and methods to show 
that evolution can make a learning of fitness landscape that  can accelerate evolution.

In this paper, we propose a model, which extends previous ones by
\citet{Valiant2006,Valiant,SatE,FV} in two aspects. First, we use hybrid  circuits involving two kinds of variables.
The first class of variables are real valued ones, they range in the interval $[0,1]$ and they can be interpreted 
as relative levels of expression phenotypic traits or morphogene concentrations 
(morphogenes can make cell differentiation and thus change phenotypes).  Other variables are Boolean and can be interpreted as genes. 
 Second, we use a threshold scheme of regulation, which inspired
by ideas of the paper by  \citet{Stern}.  All  variables are involved in the gene regulation via thresholds. 
Our results are formulated in a mathematically rigorous form   for such models, which involve  general fitness functions and population dynamics, and which are based on fundamental biological 
ideas and experimentally confirmed facts.  Namely, we assume that the multicellular organism phenotype depends on a number of phenotypical traits and these traits are completely determined
by cell patterns. These patterns, in turn, are determined by
expressions of  special genes, for example, morphogenes.  Expression of morphogenes is controlled 
by a number of other genes (genotype) and by network regulation loops that exploit threshold mechanisms 
and involve morphogenes and other genes.   Populations are large but not obligatory infinite.  
The gene drift can be taken into account if it is small. 

It is shown that our gene regulation scheme is powerful enough and allows to  realize all possible feedback mechanisms, any algorithms of gene control (see Theorem I) and can produce any dynamics.   So, such regulation scheme is powerful enough to realize any algorithms of evolution. We therefore essentially reinforce the results of \cite{Vazirani}, where it is shown that evolution can realize MWUA algorithms.

Furthermore,  the gene regulation can perform  fitness landscape recognition.   
According to our mathematical results it can be explained as follows. Consider a phenotypic trait, which is controlled, say, by many genes and
by a threshold parameter. Suppose that for a fixed threshold we need 10 mutations to obtain expression level necessary for a good adaptation.
Then in the classical mutation-selection scheme evolution uses approximately $p_{\mathrm{mut}}^{-10}$ generations.   

The canalized  evolution uses learning by gene network and it can work much faster.  Indeed, it is sufficient to use only a single mutation.
Suppose a mutation is happened, which affects the trait under consideration,  and after $T_c  \ll  p_{\mathrm{mut}}^{-1}$ generations  
some progenies of that mutant  are still viable.  
This fact on viability gives an information on the fitness landscape since it means that  that change of the trait is useful for adaptation. 
Then the gene regulation network turns on 
making a change of the threshold.  If after this change and new $T_c$ generations new progenies are still viable, then the gene regulation network 
make a new round of the threshold
variation.   

This capacity to fitness landscape recognition also explains results of
\cite{Nature} on prediction of environmental changes. 
Moreover,  gene network recognition strongly reduces the  number of mutations that are  necessary for adaptation because the gene regulatory networks 
reinforce effects of some mutations:
a single mutation can lead to a large phenotype change, the fact, which is  consistent with QTL date \citep{Zeyl}. 

So, learning based evolution can go faster since reduces need mutation number.  However,  this acceleration, 
based on stochastic learning algorithms, leads to an additional  risk connected with learning errors.  
So, evolution  reminds rates on a rugged highway:  acceleration increases risk and finally, population can come to an evolution dead end.  
Actually, a biological uncertainty principle is obtained:  to decrease the learning risk, evolution should use more
test rounds and thus a smaller regulation parameter $r_F$, however, for smaller $r_F$ we obtain greater evolution times $T_W$.

Note that evolution as a learning problem was first considered in the pioneering paper by
\citep{Valiant2006,Valiant}, where a formal Boolean circuit model  is used.  This model is essentially simpler and we think that it is less biologically realistic than 
the one considered here, since it  does not involve genetic regular networks,
which regulate Boolean circuits.  The results 
obtained by \citep{Valiant} mean that evolution may be successful for a very narrow class of circuits. Opposite to 
\citep{Valiant} we conclude that evolution may be successful  for a general class of the fitness  functions due  to  genetic regulation.  
 It is worth to note that these estimates of evolution acceleration can be done for random fitness landscapes. 
It is important fact since actually biologists know almost nothing about fitness landscapes for complex organisms \citep{Krug}.

We show that a learning by the regulatory  gene networks   also increases  canalization, i.e., phenotype stability.  
Evolution, accelerated and canalized by evolution recognition procedures, becomes  more irreversible.  This fact yields important consequences for evolution in
a varying environment.  Learning by gene networks leads to canalization,  however,  this effect  diminishes flexibility (evolvability). The fitness sensitivity with respect to mutations 
is a decreasing function of canalization level. A too strong canalization can become dangerous for population evolvability  when environment changes.

This negative effect  can be compensated by expression of other morphogenes or even a formation of new morphogenes.  
Then new structures result of old already preexisting ones. In this case evolution, as it was explained by Gould, can go by addition terminal. 
So, it is possible  that the  famous (although vague)  idea, largely discussed beginning with 19-th century, that the morphogenesis recapitulates evolution, 
can be explained by  the concept of canalized evolution. 

Another possible mechanism to avoid negative consequences of canalization for varying environments can be connected with stress. 
It is well known and proved experimentally that
under a stress (which can be induced by environmental factors, for example, temperature),  gene and proteins that make genetic buffering (for example, shaperons such as Hsp90)
change their activity that provoke mutations.   However, the most of these mutations are negative and thus the mutants obtained by this way are not well adapted.
Finally, results obtained in this paper support the idea of Waddington  that,  in a fixed environment,  evolution acceleration and phenotype  buffering are positively correlated effects. 
This correlation appears as  a result of the fitness landscape learning by gene networks.

The case of varying environments needs an additional investigation and we plan to address it in a future work.

%\section*{Figures and Tables}

%Figures and Tables should be labelled and referenced in the standard way using the \verb|\label{}| and \verb|\ref{}| commands.

%\subsection*{Sample Figure}

%Figure \ref{fig:spectrum} shows an example figure.

\section{Appendix 1}

Let us prove Theorem III.  

First we formulate an auxiliary lemma, which gives an estimate of fluctuations of the number of progenies. \vspace{0.2cm}

{\bf Lemma 1}. {\em Let $X(s,t)$ be the frequency of genotype $s$ at the moment
$t$,  $F(s)$ be the fitness of these individuals, and $\bar F(t)$ be the averaged population fitness at the moment $t$, and $\kappa=N_{pop}(t)^{-1/4}$. 
Then, if $p_{\mathrm{mut}}=0$, the frequency $X(s,t+1)$ at the moment $t+1$ lies in the interval
$$
J= [X(s,t)  F(s)\bar F(t)^{-1}(1 - \kappa),  \ X(s,t) \bar F(t)^{-1}(1 + \kappa)]
$$
with the probability $Pr_t$ satisfying estimate
\begin{equation}  \label{fluct1}
Pr_t   > P_*,
\end{equation}
where $P_*$ is defined by (\ref{Prob1}).
}

\vspace{0.2cm}

{\bf Proof}.  Recall that the number of progenies for each individual is defined by the Poisson law, 
this number has the average $F(s)/\bar F$  and the same variation. The number $N_p$ is thus a sum of identically distributed and independent random quantities.  According to the Central Limit Theorem,  for large $N_s$ the distribution of this sum is close to a normal one, with the average $N_s F(s) \bar F^{-1}$ and the same variation that gives (\ref{fluct2}) and 
completes the proof of Lemma.

{\bf Lemma 2}. {\em Let $p_{\mathrm{mut}}> 0$ and (\ref{pmutb1}) holds. Under assumptions of the previos Lemma,
we have that the frequency $X(s,t+1)$ at the moment $t+1$ lies in the interval
$$
J= [X(s,t) \eta_1, \ X(s,t) \eta_2], 
$$
where
$$
\eta_1=(F(s)\bar F(t)^{-1}(1 - \kappa)- 2p_{\mathrm{mut}}),
$$ 
$$ 
\eta_2=(F(s)\bar F(t)^{-1}(1 + \kappa)  + (N_{popmax} p_{\mathrm{mut}})(X(s,t) N_{popmin})^{-1}
$$
with the probability $\tilde Pr_t$ satisfying estimate 
\begin{equation}  \label{fluct2}
Pr_t   >  (1 - 4 P_* p_m),
\end{equation}
where $p_m$ is defined by (\ref{pmutb1}).
}

{\bf Proof}.  The proof uses the same arguments as above in the previous lemma. Let us take into account
mutations. The frequency $X(s', t)$ can be increased as a result of mutations in all the rest genotypes.
The averaged number $N_{mut}$ of such mutants is  $E_m=N_{pop}(t) p_{\mathrm{mut}} \ll 1$, where
$N_{\mathrm{pop}}(t) \in [N_{\mathrm{popmin}}, N_{popmax}]$.  Under condition    (\ref{pmutb1}) 
 the number $N_{\mathrm{mut}}$ is  a random quantity subject to the Poisson law with the mean $E_m$ and the same variation.  The probability that there will be at least one mutant is less than $2p_m$.  These arguments give the right bound of $\eta_2$ for the interval $J$.
 
 Similarly, we  can find an estimate from below for $X(s,t)$ and it gives us the left bound of $J$ that proves the lemma.   

\vspace{0.2cm}

{ \bf Proof of Theorem}.

We use Lemma 2 and  repeat the proof of  Theorem II with small modifications.  Let us introduce the quantities
$$
Q(t)=\frac{X(s, t)}{X(s',t)}.
$$
Then, by applying the lemma 2 step by step, we obtain that the following inequality
$$
Q(T_c+T_1)  >  Q(T_1)   (\frac{\eta_ 1}{\eta_2}^{T_c},  
$$
which holds with the probablity $(1-  4 P_* p_m)^{T_c}$.
Now repeating the proof of Theorem II, we obtain  the conclusion of the Theorem III.

\section{Appendix 2}

The main idea of the proof is to represent the characteristic functions $\chi_{W(z')- W(z) >0}(b)$ by the Fourier integrals as follows:
\begin{equation} \label{A1}
  \chi_{W(z^1) > W(z^0)}=\int_{0}^{+\infty} \d x  \int_{-\infty}^{+\infty}   \exp(it(W(z^1, b)  -W(z^0, b) -x)) \d t,
\end{equation}
where $i=\sqrt{-1}$.
We set $\psi(z^0, z^1)=\psi^1=W(z^1)  -W(z^0)$ and $\psi(z^*, z^1)=\psi^2=W(z^*)  -W(z^1)$.  These functions can be represented as
$$
\psi^1=\sum_{k} b_k  \psi_k^0,\quad  \psi^2= \sum_{k} b_k  \psi_k^1.
$$
Therefore,
 \begin{equation}\label{A2}
 \begin{array}{l}
 \mathrm{Pr}(Y_0 Y_1)= 
 \\
 \int_{0}^{+\infty} \int_{0}^{+\infty} \d x_1 \d x_2  \int_{-\infty}^{+\infty}  \int_{-\infty}^{+\infty} \d t_1 \d t_2  \int_{{\bf R}^{N_b}} \exp(i S ) \d \mu,
 \end{array}
\end{equation}
where $S=t_1 (\psi^1-x_1) + it_2 (\psi^2 -x_2) $.

The functions $\psi^l$ are linear in the variables $b_j$. Thus the integrals over $b_i$ have the form of typical Gaussian ones.  We compute these integrals that gives
 \begin{equation}\label{A3}
  \begin{array}{l}
 \mathrm{Pr}(Y_1 Y_0)=
 \\
  C^{-1}\int_{0}^{+\infty} \int_{0}^{+\infty} \d x_1 \d x_2  \int_{-\infty}^{+\infty}  \int_{-\infty}^{+\infty}  \exp( \Phi) \d t_1 \d t_2,
  \end{array}
\end{equation}
where
$$
\Phi=-\frac{(t_1 \psi^1 + t_2 \psi^2)^2}{4}  + it_1 x_1 + it_2 x_2.
$$
and the normalizing factor $C$ can be written in an analogous form:
\begin{equation}\label{A4}
  C=\int_{-\infty}^{+\infty} \int_{-\infty}^{+\infty}  \d x_1 \d x_2  \int_{-\infty}^{+\infty}  \int_{-\infty}^{+\infty}  \exp(\Phi) \d t_1 \d t_2.
\end{equation}
 According to (\ref{a00}) one has
$$
\Phi=-(a_{00} t_1^2  +  a_{11} t_2^2 +2 a_{01} t_1 t_2)/4  + i t_1 x_1 + it_2 x_2.
$$
Now we compute the integrals (\ref{A3}) and (\ref{A4})  over $t_1$ and then over $t_2$, which also are Gaussian ones. We obtain
$$
\begin{array}{l}
\mathrm{Pr}(Y_0 Y_1)=
\\
C^{-1}\int_0^{+\infty} \int_0^{+\infty} \exp(- \frac{x_1^2  + r^{-2} (x_2 a_{00} - x_1 a_{01})^2}{4a_{00}}) \d x_1 \d x_2,
\end{array}
$$
where  $r^2=a_{00} a_{11} -a_{01}^2$,  and
$$
C=\int_{-\infty}^{+\infty} \int_{-\infty}^{+\infty} \exp(- \frac{x_1^2  + r^{-2} (x_2 a_{00} - x_1 a_{01})^2}{4}) \d x_1 \d x_2.
$$
In the both integrals  we make the changes of variables, first  $x_1, x_2 \to x_1,   w= x_2 a_{00} - x_1 a_{01}$ and then
$x_1 =u \sqrt{a_{00}}, \   w=v \sqrt{a_{00}} r$. Finally, this procedure gives  relation (\ref{Iz}).

\section*{Acknowledgements}

%AFTER WE OBTAIN COMMENTS 
%First of all, we are grateful to the  referees that carefully studied our paper pointing out some 
%of its weaknesses and giving us some useful suggestions. 

 The second author was supported by  the grant of Russian Ministry of Education, 2012-1.2.1-12-000-1013-016. 
%and his work was initially done in the frame of the project We~1945/7-1, which was funded by {\em Deutsche Forschungsgemeinschaft}. 
 Additionally, the second author  was 
financially supported by Government of Russian Federation, Grant 074-U01.

D. Grigoriev is grateful to the grant RSF 16-11-10075 and to both MCCME and MPI f\"ur Mathematik for wonderful working conditions and inspiring atmosphere.

J. Reinitz and S. Vakulenko were supported by US NIH grant
RO1 OD010936 (formerly RO1 RR07801).

%\bibliographystyle{genetics}
%already specified by document style file
\bibliography{references2}

\end{document}